# Design, fabrication and characterisation of InGaAs/InP single-photon avalanche diode detectors

Ryan E. Warburton, Sara Pellegrini, Lionel Tan\*, Jo Shien Ng\*, Andrey Krysa\*, Kris Groom\*, John P.R. David\*, Sergio Cova\*\*, and Gerald S. Buller

*Heriot-Watt University, School of Engineering and Physical Sciences, Riccarton, Edinburgh EH14 4AS, UK*
*(R.E.Warburton@hw.ac.uk)*
*\* University of Sheffield, Department of Electronic and Electrical Engineering, Mappin Building, Mappin St., Sheffield S1 3JD, UK*
*\*\* Politecnico di Milano, Dipartimento di Elettronica e Informazione, Piazza Leonardo Da Vinci 32, 20133 Milano, Italy*

**Abstract:** This paper demonstrates the performance of planar geometry InGaAs/InP avalanche diodes, specifically designed and fabricated for Geiger-mode operation at wavelengths around 1550nm, in terms of dark count rate, single-photon detection efficiency, afterpulsing and photon-timing jitter.

## 1. Introduction

Single-photon counting and single-photon timing in the infrared spectral range, in particular at 1550nm wavelength, have become increasingly important in a number of applications such as time-resolved photoluminescence [1], optical time-domain reflectometry (OTDR) [2], eye-safe time-of-flight laser ranging [3], object recognition and imaging [4]. More recently they have been employed in quantum key distribution [5], and non-invasive testing of VLSI circuits [6]. Commercially available InGaAs/InP avalanche photodiodes (APDs) designed for use in linear multiplication mode have been experimented and investigated in Geiger mode [7,8], in order to extend the spectral range of single-photon detection beyond the limit (approximately $\lambda \sim 1000$nm) of Si single-photon avalanche diode (SPAD) detectors. These devices exhibit good single-photon detection efficiency (SPDE > 10%) and fast timing, with sub-nanosecond jitter, but they are plagued by strong afterpulsing phenomena, which severely restrict the affordable counting rate. In the present work, InGaAs/InP avalanche diode with planar geometry have been specifically designed and fabricated for developing single-photon detectors operating in Geiger-mode. This study represents a fabrication program for planar InGaAs/InP SPADs and highlights some important issues in device design.

## 2. Device structure and characterization

The SPAD design is based on a planar structure of the type originally devised for APD devices operating in linear amplification mode with separate regions of absorption grading and multiplication (SAGM) [9,10], shown in Fig 1.

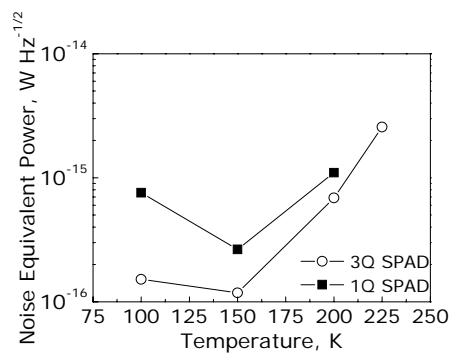

*Fig. 1* Schematic cross-section of a planar SPAD with one floating guard ring, a top p-contact to the active area and a bottom n-

*Fig. 2* Noise Equivalent Power (NEP) as a function of temperature for two designs of InGaAs/InP SPAD.

*contact to the substrate.*

Two structures were grown and characterized; one with a single quaternary (1Q-SPAD) and another with three quaternary layers (3Q-SPAD) to study the effect of grading on single-photon performance at different temperatures. For both structures we measured the single-photon detection efficiency, dark count rate (DCR), afterpulsing probability and timing jitter. The performance in terms of noise-equivalent power (NEP) is shown in fig. 2. NEP is a function of both SPDE and DCR and is calculated using the following equation.

$$NEP = \frac{h\nu}{SPDE}\sqrt{2DCR} \qquad (1)$$

In summary, we have developed SPADs for efficient single-photon detection at a wavelength of 1.55µm with a SPDE of 10% at 200K. They have dark count rates and noise-equivalent power comparable to those previously measured for commercially-available avalanche photodiodes operated in Geiger-mode [11]. Photon timing jitter at sub-nanosecond level has been measured; it is estimated that it can be further reduced through improved ohmic contacts to the detector, as well as improved packaging. We investigated the effect of the grading layer on the SPDE and fully tested the devices for robustness in temperature cycling and long operation times. The SPAD device developed still suffer from a level of afterpulsing comparable to that of the commercial APD devices previously studied. However, the new SPAD devices have reached a sufficiently high level of operating performance to allow them to be used as a basis for further analysis; a subsequent development program is planned for investigating the origin of the trap states that cause the afterpulsing phenomenon and methods for reducing their concentration.